\documentclass[smallextended,final]{svjour3}       
\smartqed  
\usepackage[francais,english]{babel}
\usepackage{natbib}
\usepackage{graphicx}
\usepackage{amssymb}
\usepackage{amsmath}
\usepackage{multirow}

\bibpunct{(}{)}{;}{a}{}{,}

\journalname{Acta Biotheoretica}

\begin{document}


\title{On the Nature and Shape of Tubulin Trails : Implications on Microtubule Self-Organization\thanks{This work was supported as an exploratory project of the Rhone-Alpine Complex System Institute IXXI.}
}
\titlerunning{Tubulin trails}        
\author{Nicolas Glade}

\institute{N. Glade \at
              AGIM Laboratory\\
	      Laboratory of AGeing Imaging and Modeling\\
	      University of Grenoble - CNRS FRE 3405\\
	      Domaine de la Merci, 38700 La Tronche, France\\
              Tel.: +33-4-56-52-00-27\\
              Fax: +33-4-76-76-88-44\\
              \email{Nicolas.Glade@ujf-grenoble.fr}
	  }

\date{Received: date / Accepted: date}

\maketitle

\selectlanguage{english}
\begin{abstract}
Microtubules, major elements of the cell skeleton are, most of the time, well organized \textit{in vivo}, but they can also show self-organizing behaviors in time and/or space in purified solutions \textit{in vitro}. Theoretical studies and models based on the concepts of collective dynamics in complex systems, reaction-diffusion processes and emergent phenomena were proposed to explain some of these behaviors. In the particular case of microtubule spatial self-organization, it has been advanced that microtubules could behave like ants, self-organizing by 'talking to each other' by way of hypothetic (because never observed) concentrated chemical trails of tubulin that are expected to be released by their disassembling ends. Deterministic models based on this idea yielded indeed like-looking spatio-temporal self-organizing behaviors.\\
\indent Nevertheless the question remains of whether microscopic tubulin trails produced by individual or bundles of several microtubules are intense enough to allow microtubule self-organization at a macroscopic level. In the present work, by simulating the diffusion of tubulin in microtubule solutions at the microscopic scale, we measure the shape and intensity of tubulin trails and discuss about the assumption of microtubule self-organization due to the production of chemical trails by disassembling microtubules. We show that the tubulin trails produced by individual microtubules or small microtubule arrays are very weak and not elongated even at very high reactive rates.\\
\indent Although the variations of concentration due to such trails are not significant compared to natural fluctuations of the concentration of tubuline in the chemical environment, the study shows that heterogeneities of biochemical composition can form due to microtubule disassembly. They could become significant when produced by numerous microtubule ends located in the same place. Their possible formation could play a role in certain conditions of reaction. In particular, it gives a mesoscopic basis to explain the collective dynamics observed in excitable microtubule solutions showing the propagation of concentration waves of microtubules at the millimeter scale, although we doubt that individual microtubules or bundles can  behave like molecular ants. 
\keywords{Microtubules \and Self-organization \and Tubulin trails \and Ant-based model \and Microscopic scale model \and Reaction-diffusion}
\end{abstract}
%
\section{Introduction}\label{introduction}
\indent In this article we aim to characterize by a model of diffusion working at the microscopic level the shape and the nature of hypothetic
heterogeneities in the free tubulin concentration profile that are expected to be produced by disassembling microtubules in \citet{Rob:BC90,Tab:ACS99,Gla:AB02,Tab:BC06}. Their existence in the form of concentrated regions of tubulin-GTP or tubulin-GDP condition the validity of certain models of microtubule self-organization\protect\footnote{In this article, the term \textit{self-organization} has a different meaning than \textit{self-assembly}. Self-assembly addresses to the process by which an individual microtubule forms spontaneously by assembly of tubulin-GTP subunits \citep{Wei:S72}. Self-assembly is of course a kind of self-organization, but here self-organization designs specifically a more macroscopic organizational level. It refers to the order that can appear from particular distributions and orientations of microtubule populations in solution.} \textit{in vitro} \citep{Pir:EJ87,Car:PNAS87,Man:S89,Tab:NB92,Pap:BC99,Tab:ACS99} based on reaction-diffusion processes \citep{Rob:BC90,Gla:AB02,Sep:PRE99}. The numerical experiments realized in this work added to experimental observations from the literature bring new elements to understand microtubule self-organization at the microscopic level.\\
\indent Here, we want to answer two fundamental questions: (i) the question of the formation of chemical trails by disassembling microtubules and (ii) the question of their relative influence during self-organizing processes. Our results are not in favor of the hypothesis of an ant-like behavior either for individual microtubules or for microtubule arrays at a microscopic level, although we think the question is still open. We however confirm the idea that large scale areas of different composition in tubulin (tubulin-GTP \textit{vs} tubulin-GDP) can form and explain large scale self-organizing phenomena in microtubule solutions, such as the temporal or spatio-temporal oscillations of microtubule concentration described in \citet{Pir:EJ87,Car:PNAS87,Man:S89}.\\
We also make a review of several experimental conditions and results that reinforce the thesis of a biomechanical scenario for spatial self-organizing solutions as proposed in \citet{Bau:JCP03,Por:PRE03,Zie:PRE04,Liu:PNAS06,Bau:BC07}, instead of a 'social insects'-based process as proposed in \citet{Rob:BC90,Tab:ACS99,Gla:AB02,Tab:BC06}.
%
\subsection{Microtubules}\label{introduction:microtubules}
Microtubules are biological supra-molecular assemblies with a micrometer scale (about 30 nm diameter and several micrometers long). They are energy-dependent reacting fibers present in most of the living cells, often in large amounts (\textit{e.g.} they represent 10\% of the total amount of proteins in neurons). In addition to actin micro-filaments and intermediate filaments, they form the cell skeleton (cytoskeleton) that gives shapes and biomechanical properties to the cells. Actin filaments and microtubules don't only constitute simple static structures ; these fibers react, grow or shrink, forming that way dynamical structures that self-adapt to the changes of cell states {--} particularly changes of energy levels {--} and to the mechanical, electrical or magnetic exogenous stimuli \citep{Vas:BR82,Tab:S94,Tab:PNAS92,Pap:PNAS00,Gla:JP02,Gla:BC05,Gla:BC06,Cra:IJDB06,Roe:P06,Gal:EBJBL06,Kro:BEC07,Col:BB07,Tab:BC07,Ing:FASEB08,Wan:BBRC08,Qia:AA08,Yan:CSB08,Sie:P09,Qia:ABBS09,Qia:IEEETAS09}. Microtubules get also involved in numerous cell functions, often constituting organelles : they form the centriole, the organizing center of the cell from which the microtubules radiate, the mitotic spindle that drives the chromosomal segregation during cell division, the elastic 'motorized arms' of cilia and flagella. They also serve as rails guiding the molecular motors that drive the active transport of phospholipidic vesicles.\\
\indent Microtubule {--} as actin filaments do {--} are able to self-assemble from their constituting bricks called tubulin heterodimers (composed of $\alpha$ and $\beta$ tubulin) \citep{Wei:S72} once warmed from 4\degre C to about 35\degre C (this range of temperatures constitutes an optimum, but
microtubules can also form at room temperature at slower rates). Tubulin heterodimers are associated either with GTP (guanosine triphosphate) or GDP (guanosine diphosphate). Both can assemble but the assembly of tubulin-GTP is more favorable than tubulin-GDP. Microtubules are usually extremely reactive and weakly stable : their ends assemble or disassemble quasi permanently without periods of stability. Their reactivity is, in first approximation, linked to the amount and the nature of the tubulin heterodimers encountered locally in the neighborhood of their ends: the reaction of assembly is more favorable when the medium is concentrated in tubulin-GTP heterodimers than in tubulin-GDP heterodimers. The reactivity of microtubules also results from the very complicated intrinsic mechanochemical dynamics that occur within their ends \citep{Van:BJ05}. Due to those
non-linear dynamics, they can show either stochastic-looking individual behaviors {--} also called dynamic instability \citep{Mit:N84} {--}, where
microtubules switch apparently in a stochastic manner between rapid phases of assembly and of disassembly, or pretty regular ones such as the well-described treadmilling process (microtubules assemble regularly at one end while disassembling at the other, causing that way the microtubules move in their direction of growth) \citep{Mar:C78}. Moreover, assembly or disassembly reactions modify locally (at the microtubule scale) or more globally
(depending on the ratio between reaction rates and molecular diffusion) the respective amounts of the different chemical species in the medium
(\textit{e.g.} tubulin-GTP, tubulin-GDP, GTP, GDP, ...). These changes can have in return an effect on the microtubule population dynamics \citep{Man:S89,Job:PRL97,Sep:PRE99,Dey:PRE05}. A synchronizing of microtubule reactivities in time and/or space can arise from this coupling between the two chemical states of tubulin, \textit{i.e.} assembled tubulin or free tubulin heterodimers, notably under the control of the regeneration
rate of tubulin-GDP into tubulin-GTP \citep{Mel:EJ88}.\\
%
%
\subsection{Behaviors of microtubules \textit{in vitro}.}\label{introduction:behavior}
\textit{In vitro} solutions that only contain purified tubulin and GTP (the energy source of the reaction) in a buffer composed of various ionic species (Mg$^{2+}$ notably) can show singular macroscopic behaviors depending on the reactive conditions (temperature, ionic concentrations, tubulin
concentration ...) (Table \ref{tab:table1}). In such solutions, one can observe behaviors from 'simple' temporal periodic oscillations of microtubule concentration within the whole solution \citep{Pir:EJ87,Car:PNAS87,Mel:EJ88,Man:EJ88,Hit:JBC90}, to spatial 'stationary' morphologies with very complicated structures \citep{Tab:ACS99}. Intermediate behaviors also exist, showing concentration waves of microtubules forming and propagating periodically throughout the solution \citep{Man:S89}. This self-organized spatial behavior looks similar to those observed in some excitable media such as the Belousov-Zhabotinskii reaction \citep{BZReaction:a,BZReaction:b,BZReaction:c,BZReaction:d}.\\
\indent Other varieties of spatial self-organization were reported in similar solutions \textit{in vitro} of purified microtubules containing also microtubule associated proteins \citep{Ned:N97,Sur:S01}. In the cells, microtubules are frequently associated with such proteins (MAPs). Most of them are molecular motors (\textit{e.g.} kinesin, dynein) that use the energy of phosphated nucleotides (GTP) to propel and move directionally at the surface of microtubules while carrying vesicles. They can also link individual microtubules between each others in such a way they form a mechanically coupled network. Some have a stabilizing effect on microtubules. To explain how these particular solutions self-organize, the authors
use numerical models based on the coupling between microtubule reactivity and mechanics biased by the reactivity and mechanical constrains induced by molecular motors \citep{Sur:S01,Ned:JCB02,Ned:COCB03}. Their numerical simulations yield very realistic behaviors suggesting that their model correctly explains how microtubules and motors self-organize. Similar works succeed in explaining the spatial organization of actin networks \citep{Rol:BJ08,Pol:JBC09,Rey:NM10}.\\
\indent The behaviors described by \citet{Pir:EJ87}, \citet{Car:PNAS87}, \citet{Mel:EJ88}, \citet{Man:EJ88,Man:S89}, \citet{Hit:JBC90} and \citet{Tab:NB92,Pap:BC99,Tab:ACS99} however imply different processes because they don't contain any molecular motor (except in some experiments in \citet{Hit:JBC90}). The cases of temporal oscillations or spatial waves of microtubule concentration are not yet ambiguous. They clearly imply the major contribution of microtubule dynamics and their coupling throughout the solution by diffusion of matter (at least tubulin heterodimers). Indeed, these alternating succession of assembling and disassembling periods were directly observed by dark field microscopy at the mesoscopic level (observation field of about dozens of microns large, largely greater than microtubule diameters) \citep{Man:S89}. Further, several models based on
reactions coupled to diffusion processes were proposed \citep{Car:PNAS87,Gla:AB02,Rob:BC90,Mar:EBJ94,Job:PRL97,Sep:PRE99} and some were able to reproduce the observed behaviors. In the article we will refer frequently to 3 different typical situations {--} (i) temporal oscillations, (ii) spatio-temporal waves, and (iii) stationary morphologies {--} first described respectively by (i) \citet{Pir:EJ87} and \citet{Car:PNAS87}, (ii) \citet{Mel:EJ88} and \citet{Man:EJ88}, and (iii) \citet{Hit:JBC90} and \citet{Tab:NB92}.\\
\indent Stationary self-organized microtubule morphologies are the more complicated to explain. Indeed, they don't clearly show as important dynamic behaviors as those described before, as to say dramatic variations of the concentration of microtubules in time (microtubule concentration is followed by turbidity measurements at 350 nm) or rapid wave propagations. In these solutions, the spatial self-organization is slow and progressive. Morphologies develop during several hours (an average of 5 hours is required for obtaining well organized morphologies, although the self-organizing process continues). After a unique 'overshoot' at the initial stages of the reaction (after about 5 minutes), the microtubule concentration stabilizes \citep{Tab:S94}. The process is obviously energy dependent (since tubulin-GTP is necessarily hydrolyzed for the microtubules to grow) \citep{Tab:N90,Tab:S94} and seems to be due to a combination of coupled diffusion and reactions \citep{Pap:BC99,Tab:ACS99}. The morphologies obtained are very reproducible, and complicated : they present alternating stripes of concentration \citep{Pap:BC99} and orientation \citep{Tab:NB92,Tab:S94} of microtubules at a macroscopic level (millimeter) and moreover show nested sub-levels of self-similar organization between the macroscopic level and the level of microtubule bundles \citep{Tab:S94}. Finally, it has been also shown that the presence of a weak external field such as gravity \citep{Tab:PNAS92,Pap:PNAS00, Gla:JP02}, magnetic fields \citep{Gla:BC05,Liu:PNAS06}, or vibrations \citep{Gla:BC06} at the early stages of the
reaction is necessary to trigger the development of these self-organized morphologies at a macroscopic level. Living cells and organisms are also
sensible to weak external fields and this often implies microtubules \citep{Cra:IJDB06}. Coherently, the understanding of the processes involved \textit{in vitro} was though to give information on the manner some external fields could act on living cells.
%
\begin{table}[H] 
\caption{\textbf{Experimental conditions that allow microtubules to self-organize spatiotemporally, temporally, or spatially.}}
\label{tab:table1}
\begin{center}
   \begin{tabular}{|c|c|c|c|}
      \hline
       & \scriptsize \em \textbf{Traveling} & \scriptsize \em \textbf{Temporal} & \scriptsize \em \textbf{Stationary}\\ 
       & \scriptsize \em \textbf{waves} &\scriptsize \em \textbf{oscillations} & \scriptsize \em \textbf{striped pattern}\\ 
       & \scriptsize \em Mandelkow et al, 1989 &\scriptsize \em (1) Carlier et al, 1987 & \scriptsize \em (4) Tabony et al, 1990 \normalsize\\ 
       &   &\scriptsize \em (2) Pirollet et al, 1987 & \scriptsize \em (5) Hitt et al, 1990 \normalsize\\ 
       &  &\scriptsize \em (3) Hitt et al, 1990 & \scriptsize \em (6) Liu et al, 2006 \normalsize\\ 
      \hline\hline
      \scriptsize MT reactivity &\scriptsize Very high & \scriptsize High &\scriptsize Limited\\
      \hline\hline
      \scriptsize Buffer &\scriptsize PIPES 0.1 M (PM) & \scriptsize (1) MES 0.1 M (MEM) &\scriptsize (4) MES 0.1 M (MEM-D$_2$O)\\
       &  & \scriptsize (2),(3) PIPES 0.1 M (PM) &\scriptsize (5),(6) PIPES 0.1 M (PM)\\
      \hline
      \scriptsize Solvent & \scriptsize H$_2$O & \scriptsize (1),(2),(3) H$_2$O & \scriptsize (4) D$_2$O\\
       &  &  & \scriptsize (5),(6) H$_2$O\\
      \hline
      \scriptsize pH & \scriptsize 6.9 & \scriptsize (1) 6.8 & \scriptsize (4) 6.75\\
       &  & \scriptsize (2) 6.75 &\scriptsize (5),(6) 6.9\\
       &  & \scriptsize (3) 6.9 &\scriptsize \\
      \hline
      \scriptsize Temperature & \scriptsize 37\degre C &\scriptsize 37\degre C &\scriptsize 37\degre C\\
      \hline
      \hline
      \scriptsize Tubulin & \scriptsize 91{--}455 $\mu$M & \scriptsize (1) 50{--}150 $\mu$M &\scriptsize (4) 54.5{--}91 $\mu$M \\
      \scriptsize (M$\simeq$ 110 kDa) &   & \scriptsize (2) 54.5 $\mu$M  &\scriptsize (5) 36 $\mu$M \\
       &   & \scriptsize (3) 136 $\mu$M  &\scriptsize (6) 54.5{--}72 $\mu$M \\
      \hline
      \scriptsize GTP & \scriptsize 6 mM & \scriptsize (1) 2 mM &\scriptsize (4) 2 mM or R.S. \\
       &   & \scriptsize (2) RS & \scriptsize (5) 1 mM  \\
       &   & \scriptsize (3) 2 mM & \scriptsize (6) 2 mM \\
      \hline
      \scriptsize MAPs &  &  &  \scriptsize (5) 1 mg/ml \\
      \hline
      \scriptsize Mg$^{2+}$ (MgCl$_2$) &   & \scriptsize (1) 12 mM &\scriptsize (4) 1 mM\\
      \hline
      \scriptsize Mg$^{2+}$ (MgSO$_4$) & \scriptsize 20 mM & \scriptsize (2) 10 mM &\scriptsize (5) 1 mM \\
       &   &  \scriptsize (3) 12 mM &\scriptsize (6) 2 mM \\
      \hline
      \hline
   \end{tabular}
\end{center}
\end{table}
\indent Once again, as for microtubule traveling waves \citep{Man:S89,Sep:PRE99}, microtubule stationary spatial self-organization has been thought to be an 'emergent' behavior; a consequence of the collective dynamics of microtubules over space and time. Moreover, the action of the weak external factors on these solutions is understood in the sense of a symmetry breaking in a 'complex system' \citep{Tab:BC06}. This approach was quite original for the reason that it was an alternative to the other plausible advanced explanations of this phenomenon of gravisensitivity. Usually, people indeed consider the biomechanics (bundling and buckling of the fibers) and/or the various static interactions between microtubule rods coupled to growth as a possible way by which they could 'feel' the gravity or other external factors and self-organize at a macroscopic level \citep{Hit:JBC90,Por:PRE03,Zie:PRE04,Bau:JCP03,Liu:PNAS06,Ing:FASEB08,Bau:BC07}.
%
\subsection{The question of microtubule spatial self-organization.}\label{introduction:SO}
\begin{figure}[H]
\begin{center}
\includegraphics[width=6cm]{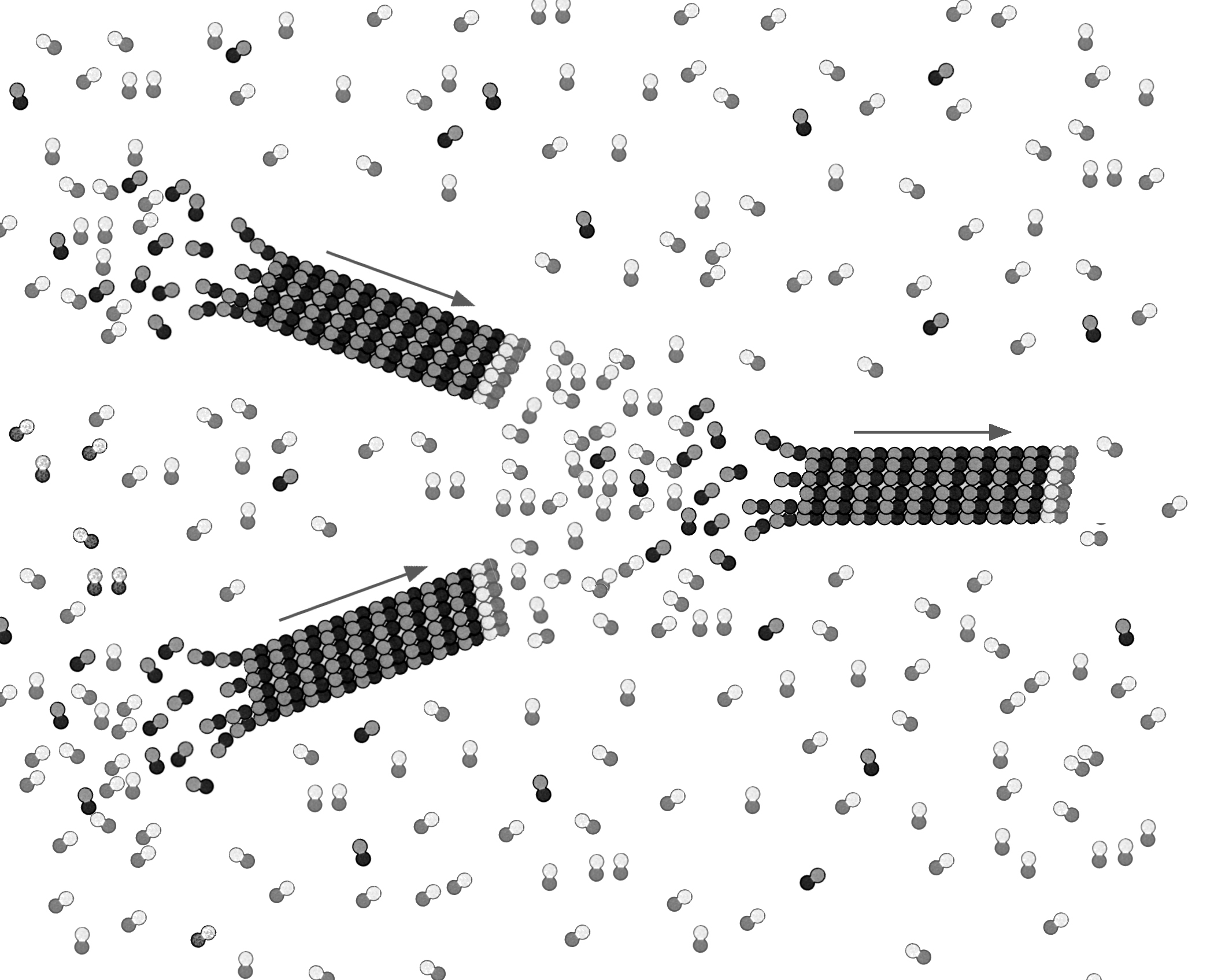}
\end{center}
\caption{ \textbf{Hypothetic self-organization scheme at the level of individual microtubules as proposed in \citet{Gla:AB02,Tab:BC06}}. In this scenario, microtubule disassembly generates concentrated trails of tubulin-GDP (in black) rapidly regenerated into tubulin-GTP (in gray). These concentrated trails are favorable media for the preferential growth (or nucleation) of neighboring (or new) microtubules. The consequence is the formation of microtubule arrays.
}
\label{fig:figure1}
\end{figure}

The following assumptions are intuitive : the chemical activity of individual microtubules or microtubule arrays should cause the formation of local
variations of concentration and composition of the chemical medium around their reacting tips. Such local heterogeneities should affect the reactivity of microtubules present in the neighborhood. This could be a way by which microtubules 'communicate'.\\
\indent It has been proposed repeatedly that the formation of such variations could influence microtubule dynamics and self-organization \citep{Tab:ACS99,Rob:BC90,Gla:AB02}. In 1990, Robert \textit{et al} published a very simple chemotactic model of microtubule self-organization where individual microtubules coordinate among themselves and self-organize, following the gradients of tubulin concentration self-produced by their own activity \citep{Rob:BC90}. Tabony published another article that reinforces the idea that microtubules behave like ants (Fig. \ref{fig:figure1}), self-organizing by collective dynamics, with emergent ant-like behaviors such as stigmergy \citep{Tab:BC06}. The author proposes that this type of physicochemical process, described at the microscopic level, is implied in living systems, explaining particularly their sensibility to weak external fields or the organizing processes that occur in the early stages of embryogenesis. Other recent studies on the effect of magnetic and electromagnetic fields or of the weightlessness on cells or microtubule solutions, works that propose biotechnological solutions based on microtubules, but also more general articles on self-organization, refer to these works on microtubule self-organization \textit{in vitro} \citep{Cra:IJDB06,Gal:EBJBL06,Yan:BP06,Roe:P06,Kro:BEC07,Col:BB07,Qia:AA08,Col:BB08,Yan:CSB08,Stra:EBM09,Myu:CAEJ09,Qia:ABBS09,Qia:IEEETAS09,Sie:P09,Qui:ASR10,Tia:MST10,Ols:IJDB10,Joh:B10,Moe:MST11,Men:ABBS11,Ayo:PRE11}. Some still take the reaction-diffusion scenario for granted and the others accept the idea that weak external fields act similarly on microtubules \textit{in vitro} and \textit{in vivo}.\\
\indent Unfortunately, the influence of the activities of microtubule ends on other microtubules by the intermediate of local variations of the tubulin concentrations {--} called tubulin trails {--} remains hypothetic since it has never been directly observed experimentally. Moreover, although the concept of reaction-diffusion is interesting and certainly significant in some microtubule solutions \textit{in vitro} at a macroscopic level
\citep{Pir:EJ87,Car:PNAS87,Mel:EJ88,Man:EJ88,Man:S89,Sep:PRE99,Dey:PRE05}, it is probably not reliable in microtubule stationary self-organizing solutions. Indeed, in the later, the influence of other effects (\textit{e.g.} mechanical or electrostatic interactions between microtubules, molecular agitation at the level of individual microtubules) is probably prevalent compared to the effect of eventual chemical trails. We do not either think that the macroscopic effects observed in microtubule stationary self-organizing solutions under to the action of external fields, can be so easily compared to those occurring in the living matter.\\
\indent Microtubule self-organization at the level of individual microtubules might be based on the existence of local trails or depletions of tubulin concentrations generated by the activity of microtubules. These local trails or depletions are indeed proposed to be the way by which microtubules 'talk to each other' \citep{Tab:ACS99,Gla:AB02,Tab:BC06}. Here we ask the question of how much assembly or disassembly reactions modify the chemical medium surrounding the ends of a single microtubule and how does this affects the reactivity of neighboring microtubules.
%
%
%
\section{Methods}\label{methods}
%
\subsection{Continuous description of matter and deterministic models of microtubule chemical trails.}\label{methods:continuous}
\begin{figure}[H]
\begin{center}
\includegraphics[width=8cm]{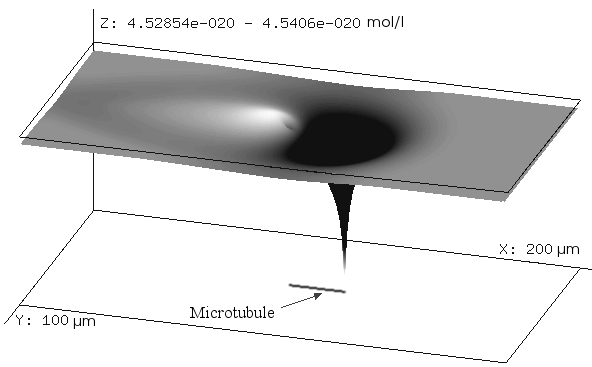}
\end{center}
\caption{\textbf{Tubulin trail in a reaction-diffusion model.} This tubulin trail is simulated using a 2D spatiotemporal differential equation system modeling a reaction-diffusion microtubule system (dimensions: $200 \times 100\ \mu m$) similar to that proposed by \citet{Gla:AB02}. The figure shows the tubulin-GTP concentration map (represented in false 3D and grey scales), in superposition of a single microtubule (above). This microtubule measures 25 $\mu$m. In this simulation, an isolated microtubule is in treadmilling motion, growing at the right end (at $21\ \mu m.min^{-1}$) and shrinking at the left one (at $18\ \mu m.min^{-1}$), in a medium containing initially $5\ mg.ml^{-1}$ {--} or $4.54\ 10^{-20} mol.\mu m^3$ {--} of tubulin-GTP. At the growing end, an 'intense' depletion of tubulin forms (the black hole). At the shrinking end, tubulin-GDP is released and rapidly converted into tubulin-GTP, producing a trail (white trail at the left of the depletion). The hole of the depletion is more intense than the peak of the trail because tubulin-GDP heterodimers have to be regenerated first into tubulin-GTP at a rate of $0.2\ s^{-1}$. During that time they diffuse all around lowering the intensity of the trail.
}
\label{fig:figure2}
\end{figure}

\indent A quantitative numerical model of formation of a depleted area of free tubulin-GTP around the growing tip of a microtubule was proposed by \citet{Odd:BJ97}. The aim of this study was to estimate if the formation of this region and its homogenization by diffusion could be a limitation for the microtubule growth reaction. The result {--} an analytical solution of a reaction-diffusion equation {--} estimated that for a microtubule growing at 7 $\mu$m.min$^{-1}$, the concentration at the tip is 89\% of the concentration far from the tip and that the concentration gradient is extending to less than 50 nm from the tip (less than 2 microtubule diameters). The finite differences implementation of partial differential equation model proposed by \citet{Gla:AB02} also predicted the formation of similar depleted areas at the growing ends of microtubules with the formation of tubulin
concentrated trails at the shrinking ends (Fig. \ref{fig:figure2}). This local phenomenon was hypothesized to be the most fundamental manner by which
microtubules communicate. It was reinforced by observations \textit{in vitro} \citep{Man:S89} and \textit{in vivo} \citep{Kea:BC99} showing that new
microtubule bundles preferentially grow in the direction and in place of shrinking ones, maintaining that way the existing microtubule bundles.
%
\indent Nevertheless, these three numerical models using or describing tubulin diffusion at microtubule ends \citep{Rob:BC90,Odd:BJ97,Gla:AB02} were simulating continuous amounts of tubulin, as to say concentrations expressed in $\mu mol.l^{-1}$, represented by floating numbers. The use of a continuous description of matter (here concentrations) is well-adapted to simulate large numbers of reactants. However, it's unadapted to simulate a small number of molecules where only discrete descriptions of matter are adapted. In real microtubular solutions, only very few tubulin-GDP heterodimers are produced by disassembling microtubules and diffuse from their reacting ends, and only few tubulin-GTP heterodimers are taken away from the medium and assembled at the microtubular tips. The use of concentrations in this case is not well adapted because, due to continuously-expressed diffusion and reactions, it can generate in the simulation tubulin amounts being fractions of individual molecules. This is exactly what is observed on Fig. \ref{fig:figure2}. The concentration map of tubulin-GTP shows small variations in space and the respective maximum and minimum are $4.5406\ 10^{-20} mol.\mu m^3$ and $4.52854\ 10^{-20} mol.\mu m^3$ which corresponds respectively to an increase of $3.6$ tubulin-GTP heterodimers (the trail) and a decrease of $69$ heterodimers (the depletion) in a volume of $1\ \mu m^3$, if we refer to the initial concentration of $4.54\ 10^{-20} mol.\mu m^3$ (equivalent to about 27340 tubulin heterodimers). The tubulin-GDP heterodimers produced at the shrinking ends are diluted in a very large amount of free tubulin heterodimers. These tubulin-GDP heterodimers can be converted by nucleotide exchange and added to the local pool of tubulin-GTP (rate $k_{reg}$ at high tubulin concentration of about $0.02 s^{-1}$ according to \citet{Mel:EJ88}). Nevertheless, the average increase is $3.6$ heterodimers in a volume of $1\ \mu m^3$, that already contains more than 27000 heterodimers. In consequence, the differences of tubulin amounts surrounding the reacting ends are very small. Moreover, the fluctuations of concentration in the other parts of the medium (far from the tips) are also very low ; they are often variations of the order of fractions of individual molecules only, which can be also viewed as very low fluctuations of the probability of presence (density) of tubulin heterodimers.\\ 
\subsection{Quantifying heterogeneity in the composition of the medium at the microscopic level.}\label{methods:quantifying}
The microscopic level is the most accurate level for describing the action of microtubule ends on the chemical medium. Working with finite numbers of
indivisible molecules instead of continuous {--} floating encoded {--} concentrations prevents the effects and the mistakes of interpretation as
described before. Moreover, having a realistic spatial representation of all the protagonist molecules present in the medium {--} free tubulin-GTP and
tubulin-GDP heterodimers and assembled heterodimers (\textit{e.g.} oligomers, nuclei, microtubules) {--} allows observing spatial encumbrance effects due for example to the elongated shape of microtubules or to the channel formed by individual microtubules, which biases the direction and speed of molecular diffusion \citep{Odd:EBJ98}.\\
\indent Yet, \citet{Van:BJ05} proposed a very accurate mechanochemical model of microtubule dynamics at the molecular level, but the consequences of
these dynamics on the local molecular environment are not studied in their article.  We are more interested here in the way tubulin dimers produced by the disassembling ends of microtubules diffuse around these ends and if they can be considered as sources of mesoscopic and anisotropic (elongated) heterogeneities of tubulin concentration and/or composition like the hypothetical tubulin trails that were advanced in \citet{Rob:BC90} and \citet{Tab:ACS99} to explain microtubule self-organization in the same way that well known trail systems like ants.\\
\indent We just mentioned that chemical trails and depletions are probably very weak. In particular, tubulin-GTP variations are hidden by the amounts present in the whole medium and, in consequence, are certainly unable to affect microtubule dynamics. Nevertheless, local heterogeneities composed of tubulin-GDP heterodimers appear and could have an influence on microtubule dynamics at a more macroscopic level. We wished to quantify such heterogeneities and to observe their survival inversely proportional to diffusion.
\indent In order to give us the better chance to obtain tubulin trails from disassembling microtubules, we caricatured the best scenario and fixed a situation where microtubule dynamics are simplified, \textit{i.e.} a situation where microtubules are only allowed to disassemble constantly at high rates (without interrupts in the disassembly such as short periods of assembly or pauses). In these simulations, microtubule disassembly consists in a constant release of tubulin-GDP, heterodimer after heterodimer, and not in the release of tubulin coiled oligomers from proto-filaments nor the release of single heterodimers described by any stochastic based process. Assembly was not permitted. The conversion of tubulin-GDP into tubulin-GTP was not either permitted so as to measure the accumulation of tubulin-GDP over simulations. This approximation is however reasonable because of the very small rate of nucleotide exchange in tubulin ($k_{reg} = 0.02 s^{-1}$ \citep{Mel:EJ88}).\\
\indent Because they are not necessary for this study, we don't describe these reactions in the article. Moreover, we don't implement any microtubular mechanics since at this scale level diluted microtubules are considered as rigid rods due to their very high persistence length (5 mm) and because we also consider that microtubule-microtubule mechanical interactions are very rare in such diluted conditions. In this context, it is allowed to simulate only the diffusion of the released tubulin-GDP heterodimer from the tip of shrinking microtubules and conclude from the observed simulated behavior.
%
\subsubsection{Simulation procedure}\label{methods:quantifying:procedure}
\begin{figure}[H]
\begin{center}
\includegraphics[width=8cm]{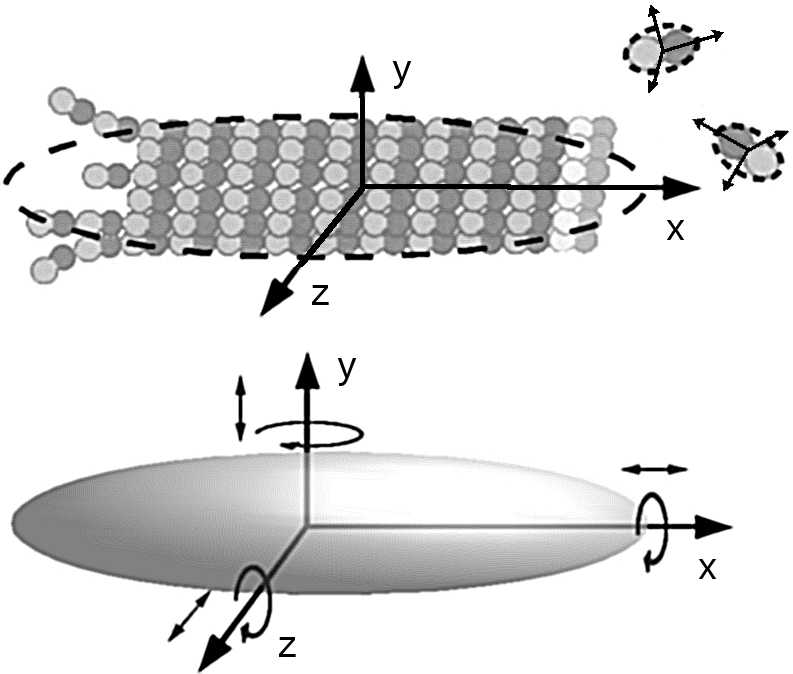}
\end{center}
\caption{\textbf{Approximation of molecular assemblies by ellipsoids.} (\textbf{Top}) an approximation of rod-like microtubules (left) and quasi-spherical tubulin heterodimers (right) by ellipsoids is shown. (\textbf{Down}) This allows correcting the translational rates of diffusion and their correlation times of the particles in solution by a shape factor calculated along each axis.
}
\label{fig:figure3}
\end{figure}

We use a microscopic representation of microtubule solutions where all tubulin heterodimers and assemblies are represented as small particles (with certain dimensions and volumes) suspended in an implicit solvent and move due to thermal agitation. Individual tubulin heterodimers are shaped as small ellipsoids (total dimension along the 3 axes : $8 \times 6.5 \times 4.6\ nm$). All reaction or diffusion events are coded in discrete time in a continuous space so as time steps can vary from nanoseconds to seconds and space steps can vary from nanoscopic levels to macroscopic levels. At each time step all tubulin heterodimers and assemblies diffuse independently and we evaluate the direction and the length of their diffusion jumps and changes of orientations. We choose this approach because we do not need a good description of the molecular kinetics for observing diffusion patterns. In the other case, when reactions are implemented, a probabilistic description of both diffusive and reactive events is more adapted. Although our implementation can also work in continuous time and semi-discrete space (not presented here) according to the Gillespie algorithm \citep{Gil:JCP76,Gil:JPC77} and as implemented in \citet{Elf:PSPIE03}, the discrete time method presented here has the advantage of being more precise in the description of diffusion at the microscopic level (microscopic or molecular diffusion).\\
\indent Preformed microtubules (radius $r=16.8\ nm$) were designed as helices with an angle step of 27.69\degre between two successive tubulin heterodimers \citep{Lan:JCB80}. As for tubulin heterodimers they can be approximated as very elongated ellipsoids.\\
\indent Usually, in numerical models of molecular diffusion, one consider isotropic diffusion of spherical particles. Given the macroscopic diffusion
constant of a particle $D$, one can calculate at each time step $dt$ the jump $dx$ of this particle in a random direction (between $0$ and $2\pi$), using a Gaussian function centered on $0$ with variance $V=2Ddt$.\\
\indent Nevertheless, many molecular assemblies in microtubule solutions have an anisotropic diffusion due to their elongated shape. For example, microtubules are elongated rods moving preferentially in the direction of their long axis, while tubulin heterodimers are quasi-spheric molecules having a quasi-isotropic diffusion. Following is the procedure used :
\begin {itemize}
\item{First, we approximate the 3D-shape of each molecule (here, tubulin heterodimers or microtubule supramolecular assemblies) by an ellipsoid (Fig. \ref{fig:figure3}). Ellipsoids are the only quadrics that can be evaluated from sets of 3D points, that give information of center and orientation of an object \citep{Li:PGMP04,Han:S06}, and for which the description of individual translational diffusion and rotational rates {--} expressed in
terms of correlation times (\textit{i.e.} the number of complete rotations during a given time) {--} along the 3 axes are described. At this spatial scale level ($<1\ \mu m^3$) and time scale level (nanoseconds to microseconds), the number of different molecular assemblies that form is limited and their length are also limited and can all be well-fitted by ellipsoids.}
\item{Using a variation of the Stoke-Einstein formula \citep{Ein:B05}, called the Perrin formula \citep{Koe:B75,Jon:JPC91}, we calculate the translational diffusion rates ($D_{_{T}k}$) and the correlation times ($D_{_{R}k}$) along each axis $a_k$ of the considered molecule. The Stoke-Einstein formula $D_k=k_bT/f_k$ is usually defined for spherical particles. Since we simulate the diffusion of elongated molecular assemblies like microtubules, we apply a shape factor $S_k$, called $S$ Perrin factor \citep{Koe:B75}, on the frictional coefficient $f_{_{T}k}=6\pi \eta rS_k$ for the translation and $f_{_{R}k}=6 \eta V_hS_k$ for the rotation, $S_k$ being equal to 1 for a spherical particle, and $V_h$ is the volume of the equivalent hydrated spherical particle. $S_k$, where $k$ designates the axis ($x$,$y$ or $z$), takes into account the shape of the ellipsoid (its 3 radii) and its nature, \textit{i.e.} oblate ((flying saucer) or prolate (cigar rod). For example, microtubules are like cigar rods, \textit{i.e.} prolate ellipsoids, whereas tubulin rings correspond more to oblate ellipsoids. There are two kinds of shape factors : one used for balancing the translational diffusion rates ($S_{_{T}k}$), the other for the correlation times ($S_{_{R}k}$). Descriptions of the algorithms for calculating the translational diffusion rates and the correlation times along the three axes are given below.}
\item{At each time step, the jump and rotation events of each molecule are obtained separately by two Gaussian functions along the 3 axes $a_k$ of the
molecule, using the respective values of $D_{_{T}k}$ and $D_{_{R}k}$} for calculating the variances of the Gaussian curves. Then, each molecule
re-orientates in the direction given by the calculated rotational rates, and moves according to its new orientation and to the calculated translational vector. Molecules diffuse randomly with the constraint that 2 molecules can not exist at the same time in the same place (their hydrated volumes can't intersect).
\end{itemize}
%
\subsubsection{Translational rates}\label{methods:quantifying:translational}
Let us consider an ellipsoid and its 3 radii $r_i$ along its 3 axes $a_i$, $i=1..3$. The translational rate $D_{_{T}k}$ along a given axis $a_k$ is only depending on the shape of the elliptic projection of the ellipsoid along this axis onto the perpendicular plane, as to say it depends on the values of the two radii $r_l,\ l \neq k$ and $r_m,\ m \neq k$. The approximative shape of a tubulin heterodimer is a short prolate ellipsoid (\textit{e.g.} a cigar) such as $r_1-r_2>r_2-r_3,\ \textrm{with}\ r_1\geq r_2\geq r_3$. Long and straight microtubules or linear oligomers of tubulin, for which the ratio $p$ of radii is very large (\textit{e.g.} $p\simeq 30$ for a $1\ \mu m$ long microtubule, about $p\simeq 8\ to\ 10$ for a tubulin ring or for a 4 heterodimers long oligomer, compared to $p\simeq 2$ for a single tubulin heterodimer), can also be well approximated by long prolate ellipsoids (\textit{e.g.} rods). Nevertheless, when the ratio between the principal radius and the secondary radius is close to 1, the projected shape is also not so far from that of an oblate ellipsoid (\textit{e.g.} a flying-saucer), as to say a disk-shaped projection. This is the case for tubulin heterodimers ($ratio=2$) or more for tubulin rings ($ratio\simeq 0.36$).\\
\indent Considering the axis of translation $a_k$, the ratio of the two perpendicular radii is $p=r_l/r_m,\ r_l>r_m$, the contribution of the prolate
shape (rod or cigar like) is $S_{_{Pro}}=\sqrt{p^2-1}/(p^3\ ln(p+\sqrt{p^2-1}))$ and that of the oblate shape (a disk) $S_{_{Obl}}=\sqrt{p^2-1}/(p^3\
arctan(\sqrt{p^2-1}))$. Ordinary shapes can not be always described as pure prolate or oblate ellipsoids. Then, the shape coefficient of any ellipsoid, $S_{_{T}k}$ for a translation along $a_k$, is defined here in first approximation as a balanced combination of both oblate and prolate contributions in such a way that $S_{_{T}k}=[S_{_{Obl}}+(p-1)\ S_{_{Pro}}]/p\quad if\quad r_l>r_m$ for any ellipsoid, and  $S_{_{T}k}=1\quad if\quad r_l=r_m$ in case of an exact disk-shaped projection. The frictional coefficient of the Stoke-Einstein formula becomes : $f_{_{T}k}=6\pi \eta S_{_{T}k} \sqrt{r_lr_m}$. For tubulin heterodimers or microtubules, $S_{_{T}1}\simeq 1$, $a_1$ being the principal axis of the molecule, because $r_2 \approx r_3$.
\subsubsection{Correlation times}\label{methods:quantifying:correlation}
Along the principal axis $a_1$ of an ellipsoid (the longer axis for a prolate ellipsoid or the shorter axis for an oblate ellipsoid), the rotation is always of prolate type. The two secondary axes of tubulin heterodimers, of straight microtubules or of tubulin rings are equal so the shape factor of rotation along their principal axis $a_1$, is $S_{_{R}1}=1$. To determine the type of rotation along the two secondary axes, we have first to determine the global shape of the fitting ellipsoid, as described before. Tubulin heterodimers or long and straight microtubules are prolate ellipsoids. Their shape factor is described as follows. Let's define $a_k$ the principal axis of the ellipsoid and $a_l$ and $a_m$ the two secondary axes. If we consider the rotation along one of the two secondary axes, $a_l$ (resp. $a_m$), the ratio of its perpendicular radii is $q=r_m/r_k$, $r_k$ being always the radius collinear to the principal axis of the ellipsoid. The shape factor that acts of the frictional coefficient of rotation along the axis $a_l$ (resp. $a_m$) is $S_{_{R}l}=4(1-q^4)/[3q^2((2-q^2)C-2)]$ with $C=ln((1+\sqrt{1-q^2})/q)/\sqrt{1-q^2}$. On the contrary, tubulin rings, for example, are oblate ellipsoids. The definition of their shape factor is the same than before with $C=2\ arctan(\sqrt{1-q^2})/\sqrt{1-q^2}$ and $q=r_k/r_m$.
\subsubsection{Tubulin-GDP amount profiles}\label{methods:quantifying:profiles}
\indent The amount profiles are measured from the geometric center of the disassembling tips of an array of 5 microtubules (see the inset, Fig. \ref{fig:figure5} showing a transverse cross section of the x axis and of 5 MTs), each of them respectively separated by 30 nm (one microtubule diameter). All microtubules disassemble simultaneously and regularly (no probabilistic events) at $20\ \mu m.min^{-1}$ (about $1.85\ ms.dimer^{-1}$) which is a quite fast disassembling rate. The amount of released tubulin-GDP heterodimers is very low (25 tubulin-GDP heterodimers are released by the 5 disassembling microtubules during the 11 ms of the simulation) and needs to be integrated in time for obtaining good average profiles. The graphic has been reconstructed by integration of the amount maps of 6 independent simulations, during 1.8 ms (\textit{i.e.} the average time separating the release of 2 tubulin-GDP heterodimers by a disassembling microtubule), between the simulation times 9.2 ms and 11.045 ms (sampling time steps are equal to $5\ 10^{-6}\ s$), along the 3 axis (\textit{i.e.} a total of 6642 profiles).
\subsubsection{Continuous model fitting}\label{methods:quantifying:fitting}
\indent To estimate the macroscopic diffusion rates from the tubulin-GDP amount profiles, we reproduced the numerical experiment in a continuous model of diffusion, in which 5 tubulin-GDP heterodimers are produced every 1.844 ms (5 times between 0 and 9.22 ms) at the position $x=0$. This is obtained by summing 5 times the equation of diffusion $(N_0 / (2\sqrt{\pi D t})).exp(-x^2 /(4 D t)) $ from a source, each equation shifted in time by 1.844 ms. The integration of the signal every $5\ 10^{-6}s$ between 9.22 ms and 11.064 ms is a sum over this period and with the same time step of this sum of time-shifted diffusion equations. Two continuous models are considered : the one with only one diffusion parameter $D$ corresponds to a continuous uniform diffusion model ; the other has two diffusion parameters $D_{InBundle}$ and $D_{OutOfBundle}$ which correspond respectively to the diffusion of tubulin-GDP heterodimers within the microtubule bundle (area approximately comprised between -0.1 and 0.1 $\mu m$) and the diffusion outside of the bundle. A large range of diffusion parameters is explored so as to find the optimal diffusion parameters that allow to fit the profiles of integrated tubulin-GDP amounts. We also allowed to fit the profiles shifted to larger values in the limit of their standard deviation. The lowest sum square error between the recorded profiles and the continuous models indicates the optimal fit, \textit{i.e.} the optimal diffusion parameters.
%
\section{Results}\label{results}
%
\subsection{Values of the diffusion rates.}\label{results:values}
\indent For individual tubulin heterodimers, we obtained the following translation rates and correlation times (table \ref{tab:table2}) for water conditions as for \textit{in vitro} solutions of microtubules (viscosity $\eta_{Water,\ 37\textrm{\degre} C}=6.915.10^{-4}\ N.m.s^{-2}$) or for
cytoplasmic conditions as in living cells (viscosity $\eta_{Cytoplasm,\ 37\textrm{\degre} C}=5.6.10^{-3}\ N.m.s^{-2}$ \citep{Sal:JCB84}).\\
%
\begin{table}[H]
\caption{\textbf{Diffusion rates and correlation times of tubulin heterodimers.} These diffusion parameters calculated for tubulin heterodimer elliptic approximate take into account the viscosity of water or of the cytoplasm at $37\textrm{\degre}C$.}
\label{tab:table2}
   \begin{center}
	\begin{tabular}{|l l||r@{.}l@{ }l|r@{.}l@{ }l|}
	\hline 
	\multicolumn{8}{|c|}{\textbf{\em Tubulin heterodimer}}\\
	\multicolumn{8}{|l|}{\small \ Translational diffusion rates $D_T$ in $m^2.s^{-1}$}\\
	\multicolumn{8}{|l|}{\small \ Correlation times (rotations) $D_R$ in $s^{-1}$}\\
	\textbf{\em \small Axis (cf. Fig. \ref{fig:figure3})} & & \multicolumn{3}{c}{\textbf{\em \small Water}} & \multicolumn{3}{c|}{\textbf{\em \small Cytoplasm}}\\
	\hline 
	\hline 
	\multirow{2}{*}{\small\em x}	&\small $D_T$ &\small 65&\small3&\small$10^{-12}$ &\small 8&\small07&\small$10^{-12}$\\
				&\small $D_R$ &\small 22&\small1&\small$10^{6}$ &\small 2&\small73&\small$10^{6}$\\
	\hline 
	\multirow{2}{*}{\small\em y}	&\small $D_T$ &\small 52&\small6&\small$10^{-12}$ &\small 6&\small62&\small$10^{-12}$\\
				&\small $D_R$ &\small 6&\small54&\small$10^{6}$ &\small 0&\small808&\small$10^{6}$\\
	\hline 
	\multirow{2}{*}{\small\em z}	&\small $D_T$ &\small 53&\small6&\small$10^{-12}$ &\small 6&\small49&\small$10^{-12}$\\
				&\small $D_R$ &\small 6&\small16&\small$10^{6}$ &\small 0&\small76&\small$10^{6}$\\
	\hline 
	\end{tabular}
   \end{center}
\end{table}
\indent The same technique was applied to determine the translational diffusion rates and correlation times of tubulin assemblies, in particular microtubules. As an example we estimated their values for a microtubule of $1\ \mu m$ long. Values are given in table \ref{tab:table3}.
\begin{table}[H]
\caption{\textbf{Diffusion rates and correlation times of a microtubule of $1\ \mu m$ long.} These diffusion parameters calculated for the elliptic approximate of a $1\ \mu m$ long microtubule take into account the viscosity of water or of the cytoplasm at $37\textrm{\degre}C$. The microtubule is a cylinder so its axes $y$ and $z$ are equivalent.}
\label{tab:table3}
   \begin{center}
	\begin{tabular}{|l l||r@{.}l@{ }l|r@{.}l@{ }l|}
	\hline 
	\multicolumn{8}{|c|}{\textbf{\em Microtubule of $1 \mu m$ long}}\\
	\multicolumn{8}{|l|}{\small \qquad Translational diffusion rates $D_T$ in $m^2.s^{-1}$}\\
	\multicolumn{8}{|l|}{\small \qquad Correlation times (rotations) $D_R$ in $s^{-1}$}\\
	\textbf{\em \small Axis (cf. Fig. \ref{fig:figure3})} & & \multicolumn{3}{c}{\textbf{\em \small Water}} & \multicolumn{3}{c|}{\textbf{\em \small Cytoplasm}}\\
	\hline 
	\hline 
	\multirow{2}{*}{\small\em x}	&\small $D_T$ &\small 19&\small8&\small$10^{-12}$ &\small 2&\small44&\small$10^{-12}$\\
				&\small $D_R$ &\small 3360&\small0& &\small 291&\small0&\\
	\hline 
	\multirow{2}{*}{\small\em y or z}	&\small $D_T$ &\small 0&\small197&\small$10^{-12}$ &\small 0&\small0244&\small$10^{-12}$\\
				&\small $D_R$ &\small 5&\small0& &\small 0&\small618&\\
	\hline 
	\end{tabular}
   \end{center}
\end{table}
\indent From these simulations, we verified the value of the macroscopic diffusion of the population of tubulin heterodimers measured experimentally in \citet{Sal:JCB84}. We followed the diffusion of about 300 tubulin-GDP heterodimers in solutions containing fixed concentrations of 10 $mg.ml^{-1}$ of tubulin-GTP during several seconds of real time. During a simulation, the pathway of each molecule and its distance from its initial position are recorded. Toric boundary conditions are used for tubulin-GTP heterodimers to maintain their initial density in the sample, whereas boundaries are permeable for tubulin-GDP heterodimers, to obtain their correct amount profiles. The estimated values obtained here were perfectly consistent with the values measured experimentally : $D_{tub,\ cytoplasm}=5.9\ 10^{-12}\ m^{2}.s^{-1}$ in the cytoplasm and $D_{tub,\ water}=48\ 10^{-12}\ m^{2}.s^{-1}$ in water. We also carried simulations of populations of non-interacting short microtubules ($1\ \mu m$ long) in diluted solution and yield the following macroscopic diffusion constants : $D_{MT\ 1\ \mu m,\ cytoplasm}=0.55\ 10^{-12}\ m^{2}.s^{-1}$ in the cytoplasm and $D_{MT\ 1\ \mu m,\ water}=5.6\ 10^{-12}\ m^{2}.s^{-1}$ in water.
%
%
\subsection{Formation of tubulin-GDP heterogeneities in the medium.}\label{results:heterogeneities}
\begin{figure}[H]
\begin{center}
\includegraphics[width=8cm]{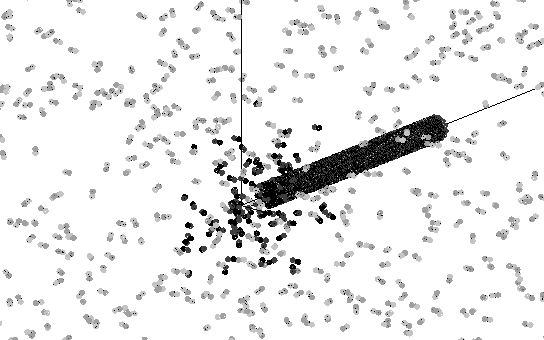}
\caption{\textbf{Concentrated area of tubulin-GDP forming at the shrinking end of a fast disassembling microtubule.} Tubulin-GTP heterodimers are displayed in gray while Tubulin-GDP is in black. In this simulation, the microtubule disassembles 100 x faster (at 2000 $\mu$m.min$^{-1}$) than a normal microtubule.}
\label{fig:figure4}
\end{center}
\end{figure}

\begin{figure}[H]
\begin{center}
\includegraphics[width=0.99\textwidth]{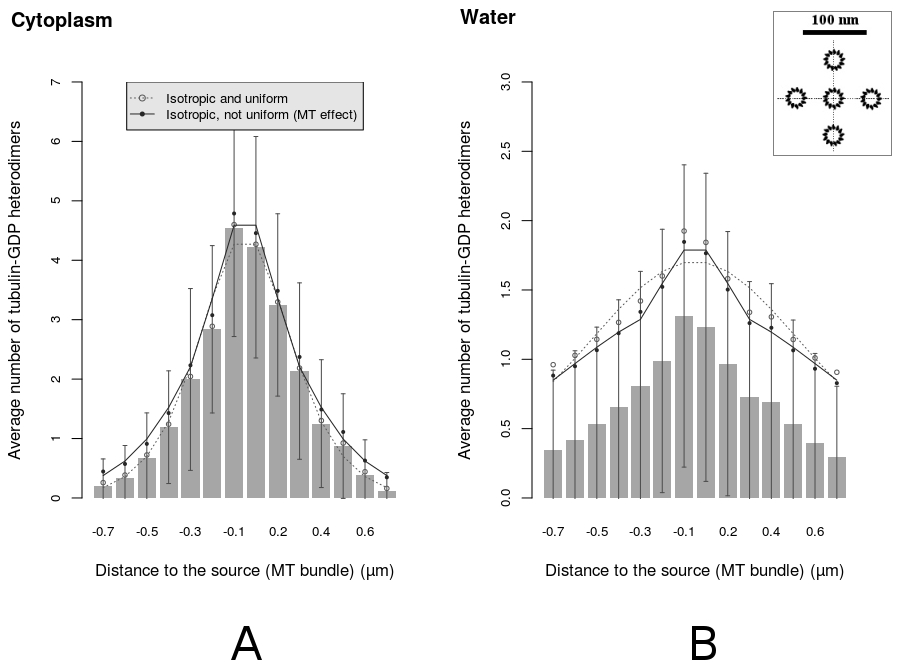}
\caption{\textbf{Profiles of tubulin-GDP amounts around the tips of 5 disassembling microtubules.} (\textbf{A}) Profiles produced by 5 motionless microtubules (the inset shows a transverse cross section of the x axis and of 5 MTs), in cytoplasmic conditions. The macroscopic diffusion rate of individual tubulin heterodimers measured from their trajectories corresponds to that measured in the cytoplasm (5.9 10$^{-12}$ m$^{2}$.s$^{-1}$) \citep{Sal:JCB84}. (\textbf{B}) The same in water at 37\degre C. The average value of the macroscopic diffusion constant measured from their trajectories (4.9 10$^{-11}$ m$^{2}$.s$^{-1}$) is about 8 times larger than in the cytoplasm, as measured in water by \citet{Sal:JCB84}. The resulting amount profile of tubulin is weaker but detectable with 5 disassembling microtubules. In both graphics, the fits produced by continuous models based on a uniform diffusion or on a microtubule dependent diffusion are drawn. The best fit of the profile shapes can be obtained after shifting the profiles to bigger values within the standard deviation of the profiles.}
\label{fig:figure5}
\end{center}
\end{figure}

To test the formation of tubulin trails, we began by positioning a single microtubule in the middle of a simulated sample of cubic shape of 1.4 $\mu$m side, oriented along the X axis, in a medium containing 10 $\mu$M tubulin-GTP (5400 heterodimers.$\mu$m$^{-3}$)\footnote{Simulations have shown that tubulin concentrations of 100 $\mu$M or the presence of numerous microtubules distributed in the sample doesn't change the macroscopic diffusion rate of free tubulin or small molecules (results not shown). \citet{Liu:PNAS06} also mention that the packing geometry of microtubular bundles doesn't avoid tubulin diffusion within the bundles. However, our simulations show that increasing concentrations of microtubules modify locally the diffusion rates of tubulin heterodimers.}. The microtubule was not allowed to diffuse and stayed located on the $x$ axis. In a first simulation, the microtubule was disassembling 100 x faster than a real microtubule (normally shrinking at a maximum rate of about 20 $\mu$m.min$^{-1}$), as to say at a rate equivalent to 2000 $\mu$m.min$^{-1}$. We realized this numerical experiment in unrealistic conditions so as to obtain approximately what was expected in \citet{Tab:ACS99,Gla:AB02,Tab:BC06,Rob:BC90}. In this case indeed, an eye-observable concentrated area formed around the shrinking end (Fig. \ref{fig:figure4}).
\indent At realistic rates of disassembly (20 $\mu$m.min$^{-1}$) however, the variation of tubulin-GDP amount around the tip was very hard to detect, in particular in water. Indeed, molecular diffusion is a very fast homogenizing process compared to the microtubule disassembly process that creates heterogeneities. During the release of one tubulin heterodimer in the cytoplasm by a microtubule, the last released tubulin-GDP heterodimer has sufficient time (about 1.85 ms) to explore an average sphere of 150 nm of radius (0.5 $\mu$m in water). In consequence, during microtubule disassembly {--} or assembly \citep{Odd:BJ97} {--} the solution is rapidly homogeneous at the micrometer scale ($<1-2\ \mu m$). Nevertheless, although it's not intense, the very weak gradient of tubulin-GDP heterodimers can be measured in the simulations (Fig. \ref{fig:figure5}). The gradient shape obtained after integration of numerous simulations is close to a sum of Gaussian distributions (solutions of the diffusion equation that started at different times, \textit{i.e.} when tubulin-GDP heterodimers are released by microtubules) and roughly speaking, the tubulin-GDP heterodimers are nearest to
the tip than very far away.\\
\indent This can be better observed by allowing several microtubules to disassemble in the same place. We simulated 5 aligned and parallel microtubules disassembling at 20 $\mu m.min ^{-1}$ during 11 ms. A simulation time of 11 ms is a compromise so as to have a sufficient amount of tubulin-GDP released by the 5 disassembling microtubules and to have a limited diffusion area (less than 2 $µm$) to avoid simulating huge volumes. This time, the formation of a gradient of tubulin-GDP was eye-observable and measurable. After 11 ms of reaction in cytoplasmic conditions, tubulin-GDP heterodimers are observed at a distance of about 0.7 $\mu$m from the tips of the microtubules from where they were released and the gradient has an average maximum value of 4.5$\pm$1.8 heterodimers at the disassembling microtubule tips (Fig. \ref{fig:figure5} A). In water, the intensity of this gradient is weaker. After 11 ms, the gradient of tubulin-GDP extends more rapidly until a distance (observable tubulin-GDP heterodimers) of about 1.5 $\mu$m from the tips of the microtubules. It has an average maximum value of 1.31$\pm$1.08 heterodimers at the tips of the microtubules (Fig. \ref{fig:figure5} B).
%
As seen in table \ref{tab:table4}, the values of diffusion estimated from the average recorded pathways of tubulin-GDP, from the uniform diffusion continuous model and microtubule dependent diffusion continuous model are comparable. However, the values given in table \ref{tab:table4} and the model fits shown in Fig. \ref{fig:figure5} clearly confirm that only microtubule concentration dependent models can describe correctly the tubulin-GDP amount profiles shown in Fig. \ref{fig:figure5}. This result disagrees with \citet{Liu:PNAS06} : the packing geometry of microtubular bundles seems tp avoid tubulin diffusion within the bundles.
%
\begin{table}[H] 
\caption{\textbf{Macroscopic diffusion estimated from the profiles shown in figure \ref{fig:figure5}.} Diffusion rates are estimated from the average recorded pathways of tubulin-GDP heterodimers, or from continuous models based on a uniform diffusion or a microtubule dependent diffusion. In the latter, diffusion of tubulin-GDP heterodimers is different inside and outside the microtubule bundle. Sum square errors (SSE) between the model fit and the profiles are indicated.}
\label{tab:table4}
\begin{center}
   \begin{tabular}{|c|c|c|}
      \hline
      \small \em \textbf{Medium} & \small \em \textbf{Quantifying} & \small \em \textbf{Diffusion}\\ 
	& \small \em \textbf{method}& \small \em ($m^2.s^{-1})$\\ 
	& & \small \em (and SSE of the fitting model)\\ 
      \hline\hline
      \small cytoplasm & \small Trajectory records &\small $5.9\ 10^{-12}$\\
      \hline
      \small cytoplasm & \small Uniform &\small $5.35\ 10^{-12}$\\
       & \small diffusion model &\small (SSE: $0.44$)\\
      \hline
      \small cytoplasm  & \small MT dependent &\small $2.57\ 10^{-12}$ (inside bundle) \\
        & \small diffusion model &\small $8.97\ 10^{-12}$ (outside bundle)\\
        &  &\small (SSE : $0.22$)\\
      \hline\hline
      \small water & \small Trajectory records &\small $4.9\ 10^{-11}$\\
      \hline 
      \small water & \small Uniform &\small $3.79\ 10^{-11}$\\
       &  \small diffusion model &\small (SSE: $0.15$)\\
      \hline
      \small water  & \small MT dependent &\small $2.25\ 10^{-11}$ (inside bundle) \\
        & \small diffusion model &\small $5.69\ 10^{-11}$ (outside bundle)\\
        &  &\small (SSE : $0.015$)\\
      \hline
   \end{tabular}
\end{center}
\end{table}
\indent Microtubules, by their assembling or disassembling activity can produce local chemical heterogeneities. Unfortunately, even if dozens of disassembling microtubules are located in the same place, the increase of total tubulin (tubulin-GTP and tubulin-GDP) concentration is undetectable: the variation of tubulin-GDP amount is indeed much weaker than the natural fluctuations of total tubulin amount (there are about 1000 tubulin-GTP heterodimers per released tubulin-GDP heterodimer). These heterogeneities are not concentrated areas but more areas in which the composition is changed.\\
\indent For individual microtubules, the heterogeneities are very weak and extended in space. In consequence, particularly in \textit{in vitro} solutions (viscosity of water), it is very improbable that an individual microtubule can influence another one in its neighborhood. The effect can become significant when produced by a group of synchronously reacting microtubules. To produce an intense composition or concentration heterogeneity, there are two possible scenarios: (1) in a solution of randomly distributed microtubules, several dense nodes of microtubule ends naturally exist that can form initial local composition (or concentration) heterogeneities; (2) microtubule arrays (bundles) form locally by another mechanism (for example by static interactions \citep{Liu:PNAS06,Bau:BC07}). Then, in both cases, the formation of heterogeneity nodes can potentially provoke the nucleation, the recruitment or the inhibition of microtubules as proposed in \citet{Gla:AB02, Rob:BC90}. Moreover, because of the rapid diffusion of tubulin heterodimers in comparison to the reactivity of microtubules, tubulin 'trails' are not directional as proposed in the reaction-diffusion ant-based model presented in \citet{Tab:ACS99,Gla:AB02,Tab:BC06}. As their shapes seem isotropic, the trails produced by microtubules might be inefficient to serve as guiding chemical rails for other growing microtubules.\\
\\
\indent Other effects could reinforce a little the intensity of tubulin-GDP concentrated areas. First, we observed a effect of microtubule density on local macroscopic diffusion : diffusion of tubulin heterodimers is about 3 times slower within bundles. Microtubule bundles should then contain concentrated stocks of tubulin-GDP, in particular in bigger bundles. Moreover, microtubules can release individual tubulin-GDP dimers but also oligomers of several assembled tubulin heterodimers. As calculated by our algorithm, their diffusion rate is about 2 times slower than that of free tubulin (for example, a straight oligomer of 10 subunits have the following diffusion rates $D_{oligo\ 10\ subunits,\ cytoplasm}=2.8\ 10^{-12}\ m^{2}.s^{-1}$ and $D_{oligo\ 10\ subunits,\ water}=23\ 10^{-12}\ m^{2}.s^{-1}$ compared to $D_{tub,\ cytoplasm}=5.9\ 10^{-12}\ m^{2}.s^{-1}$ and $D_{tub,\ water}=48\ 10^{-12}\ m^{2}.s^{-1}$). This could help a little to maintain a little the free tubulin-GDP more concentrated in the neighborhood of the tip.
%
\subsection{Effect of microtubule diffusion.}\label{results:MTdiffusion}
\begin{figure}[H]
\begin{center}
\includegraphics[width=0.99\textwidth]{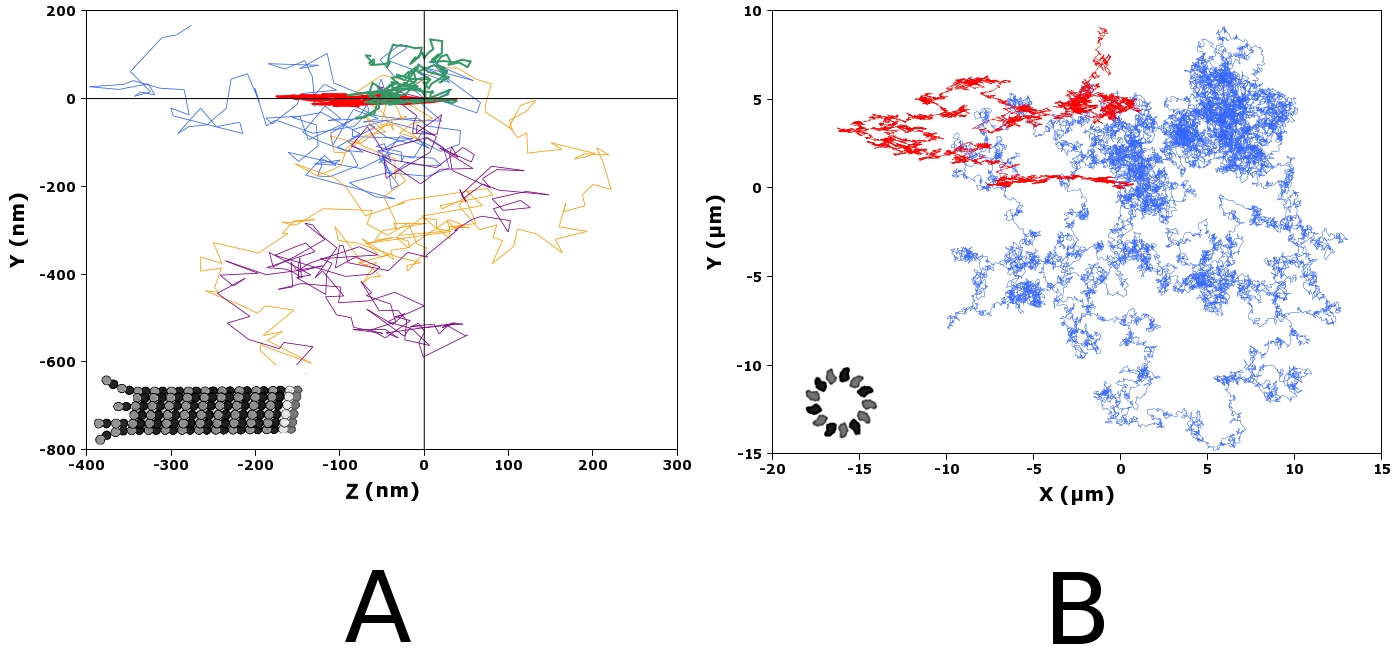}
\caption{\textbf{Time-record trajectories of microtubules and tubulin heterodimers.} Projections, from the 3D space to the YZ (A) and XY (B) planes, of 3D trajectories of 3 individual tubulin heterodimers (orange, blue and violet) and 2 microtubules of 0.1 $\mu m$ (green) and 1 $\mu m$ (red). Microtubules are initially oriented along the Z axis (perpendicularly to the plane XY). At the left bottom of each trajectory record, a scheme indicates the initial orientation of microtubules. The trajectories are relative to their starting point and were recorded during 11 ms (B) and 3 s (B). Sampling time steps are equal to $5\ 10^{-6}\ s$.}
\label{fig:figure6}
\end{center}
\end{figure}

\begin{figure}[H]
\begin{center}
\includegraphics[width=0.99\textwidth]{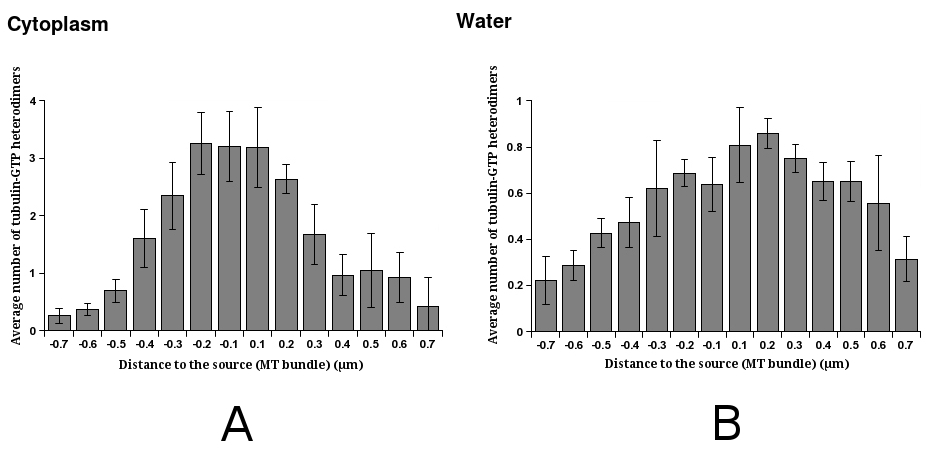}
\caption{\textbf{Profiles of tubulin-GDP amounts released by an array of 5 disassembling and diffusing microtubules.} Microtubules (of initial length equal to 0.5 $\mu m$ long) are allowed to diffuse in water (A) or in the cytoplasm (B) during the simulation (11 ms). Both are measured as described in fig \ref{fig:figure5} from the initial geometric center of the disassembling tips of microtubules (defined at the beginning of the simulation).}
\label{fig:figure7}
\end{center}
\end{figure}

In the simulations described in the previous chapter, microtubules were assumed to be motionless (cf. Fig. \ref{fig:figure5}). In real solutions, they can diffuse when sufficiently small, not blocked by other interacting microtubules and not bounded to the walls of the sample. Their diffusion is rapidly limited by their size, but the diffusion of microtubules of about 0.1 to 1 $\mu m$ long or more is not negligible. While tubulin particles have an isotropic diffusion, in agreement with \citet{Han:S06}, microtubules have a 'short time anisotropic' diffusion and a 'long-time isotropic diffusion' due to their anisotropy of shape (see Fig. \ref{fig:figure6} showing records of individual trajectories of tubulin dimers or of microtubules).
Their diffusive motion will then spread more the patterns of tubulin-GDP diffusion\footnote{The diffusion constant that describes the spreading of this pattern (the trail) results from coupled tubulin-GDP and microtubule diffusion but also depends on microtubule disassembling rate.} obtained in Fig. \ref{fig:figure5} by changing the places where tubulin-GDP is released (Fig. \ref{fig:figure7}). We did not estimate the macroscopic diffusion rates in this condition because we would need a much more complicated continuous model for, in which the microtubule dependent diffusion area would change over time. 
%
\indent In these simulations, microtubules were not grouped in the form of bundles. They could interact and collide but had no cohesion. In real solutions, microtubules can form bundles densely populated where microtubules stay grouped, behaving more like in Fig. \ref{fig:figure5}. As seen before, this may reduce this spreading effect due to microtubular diffusion.
%
\section{Discussion}\label{discussion}
\indent In this article, we asked the question of whether two (or more) microtubules close together in a solution of tubulin can communicate by way of
chemical interactions as observed in collective systems like ant colonies and whether this can explain microtubule spatial self-organization from microscopic to macroscopic levels.\\
\indent Foraging ants normally walk randomly but they can let behind them concentrated trails of chemicals called pheromones. This communication way is a manner by which is recorded the pathway of ants between a source of food and their nest once food has been found. When neighboring ants cross these trails, the probability they use the signalized pathway is greater than continuing to walk randomly. This communication way is very efficient because the chemicals released by these insects are very concentrated locally, even after a long period of diffusion (and degradation), because the dispersion area by diffusion around the original trail is of the same order as that of the size of the agents (the ants), because a trail of
pheromones is an isolated signal, not so much diluted in an environment full of equivalent pheromones, and because, due to these reasons and because social insects are sensible to weak concentrations of pheromones, neighboring ants use efficiently these trails, release pheromones within the same trail, and the signal reinforces.\\
\indent On the contrary, the present numerical simulations have shown that the tubulin trails are of very low intensity, that the magnitude of diffusion of these 'trails' is largely greater than the size of microtubule ends, that the signal of the trails is completely diluted in an environment full of tubulin heterodimers, and that in diluted solutions microtubules continue to diffuse randomly, don't stay growing within the trails and do not work at reinforcing them. This is what our numerical simulations predict until the tubulin trails will possibly be observed experimentally or not.
Moreover, even if dozens of microtubules produce a common trail, this would be certainly too weak to change locally {--} at a microscopic scale {--}
the reactivity of neighboring individual microtubules. It results that the explanation based on tubulin concentrated trails that constitutes the basis of microtubule self-organization at the individual level in stationary self-organized microtubule solutions (\citet{Rob:BC90,Tab:ACS99,Gla:AB02,Tab:BC06}) is probably unfit.\\
\indent Self-organization as observed in \citet{Tab:NB92,Tab:N90,Tab:S94,Pap:BC99,Tab:ACS99} is, of course, a kind of dissipative process : one can not deny that the reactivity of microtubules is implied during the development of the self-organizing pattern and its maintaining. However the description of that system is incomplete because it only refers to a reaction-diffusion system expressed in terms of Turing-type processes \citep{Tab:S94} and collective dynamics in the meaning of social insects \citep{Rob:BC90,Tab:ACS99,Gla:AB02,Tab:BC06}. In these microtubule solutions, there are reactive processes and transport of matter, at least by molecular diffusion processes, but other processes such as biomechanical or electrostatic interactions should also have very strong contributions. The question now is to know how strong the coupling of reaction and diffusion processes is compared to the influence of static ones. On this point, our simulations can not yet answer. It is indeed plausible that static processes play the most important role in the microtubule self-organizing process and that the existing tubulin-GDP trails (estimated in this study) reinforce it or even biased it to cause cause complex morphologies.\\ 
\indent In their paper, \citet{Liu:PNAS06} analyzed how the microtubular self-organized stripes formed. They suggest that microtubules are packed into
bundles and buckle due to their growth. In this scheme, the self-organized microtubule stripes form by this mechanism from previously aligned microtubule bundles that are initiated either by static magnetic fields or convective flow (shearing). This mechanism is similar to those proposed by \citet{Por:PRE03,Bau:JCP03,Zie:PRE04,Bau:BC07}, that couple microtubule growth to microscopic self-ordering due to nematic ordering, the macroscopic self-ordering being biased by gravity.\\
\indent A kinetic experiment of neutron scattering on a microtubular self-organizing structure examined through a horizontal slit (dimensions : 4 x
0.5 mm), described in \citet{Tab:S94} (see Fig. 4 of this article) and obtained in microtubule stationary self-organized morphologies, suggested that microtubules disassemble and reassemble during the organizing process. Because of this, microtubules are though reorient and form progressively the patterns of concentration and orientation. Microtubules were oriented in a preferential direction within the first 2 hours, then the time dependence of microtubular scattering intensity showed a decrease at about 2 hours, indicating a loss of any preferential orientation, and then increased
again indicating a reorientation along the opposite direction. However, despite of the interpretation given by the author, it does not indicates inevitably a 'partial disassembly-reassembly-disassembly' process, nor it gives information on what really occurs at the microscopic level. This interpretation would be right granted that microtubule packs can't move in the solution. It only indicates that microtubule orientation changes and that, at one moment, their orientation seems isotropic in the field of observation. This result could be interpreted differently, in agreement with the bundling and nest buckling process mentioned before : packs or single bundles of aligned microtubules buckle due to microtubule growth. This causes changes of the orientation of the bundled microtubules and also causes their mechanically-driven travel (in opposition to the treadmilling travel of microtubules due to their reactivity) across the solution and the observation window.\\
\\
At the beginning of the article, we mentioned that different mechanisms could be involved in the different varieties of microtubular self-organizing behaviors. On the one hand we have chemical solutions in which microtubules are very dynamic (as observed in time series of the concentration of assembled tubulin measured by spectrophotometry at 350 nm) and that clearly imply reaction coupled to diffusion processes ensuring synchronizing phenomena in microtubule populations \citep{Pir:EJ87,Car:PNAS87,Man:S89} {--} although it does not imply that static processes of ordering are absent. On the other hand, spatially self-organizing solutions of microtubules \citep{Tab:N90,Hit:JBC90,Tab:NB92,Tab:S94,Pap:BC99,Tab:ACS99,Liu:PNAS06} are less dynamic (only one overshoot of microtubule concentration is observed) and probably involve other contributing mechanisms such as biomechanical and/or nematic ordering \citep{Hit:JBC90,Por:PRE03,Bau:JCP03,Zie:PRE04,Liu:PNAS06}. However, the reactants are very similar from one of these \textit{in vitro} solutions to another. The major differences (Table \ref{tab:table1}) are the concentration of magnesium ions (20 mM in Mandelkow solutions, about 10 mM in those of Carlier, Pirollet and Hitt ones, and only 1 mM in Tabony and Hitt solutions), the use of deuterium oxide instead of water only in Tabony solutions, and the presence of stabilizing MAPs in Hitt solutions. Magnesium ions are known to increase the reactivity of microtubules : they promote microtubule assembly and disassembly \citep{Fly:PRE96}. On the contrary, deuterium oxide has been shown to stabilize microtubules\footnote{D$_2$O also prevents tubulin proteins against denaturation, this allowing to realize long term experiments} \citep{Cha:BC99} : it suppresses the dynamic instability and the treadmilling behavior of microtubules \citep{Pan:BC00} but stimulates the nucleation of new microtubules from free tubulin heterodimers \citep{Ito:BBA84}. In microtubule solutions where stationary self-organized morphologies appear \citep{Tab:NB92} everything tends to produce and maintain numerous, very stable and probably very long microtubules whereas it is exactly the contrary in the case of very reactive solutions that favor the formation of temporal oscillations of of traveling waves of microtubule concentration.\\
\indent This re-examination of experiments on microtubules over the last 20 years gives arguments, which add to the numerical results presented
here, against a 'molecular ants'-based scheme and against a reaction-diffusion scheme for the stationary well organized spatial morphologies described in \citet{Tab:N90,Hit:JBC90,Tab:NB92,Tab:S94,Pap:BC99,Tab:ACS99,Liu:PNAS06}. On the contrary, in this case, the thesis of a self-organizing mechanism based on the biomechanics of individual growing microtubules and bundles appears more reasonable.\\
\indent Nevertheless, in microtubule solutions showing periodic temporal variations or spatio-temporal variations (propagating waves) of the concentration of microtubules, the contribution of tubulin-GDP 'clouds' produced by individual microtubules or by bundles of numerous microtubules can't be ignored. In these solutions, microtubules are synchronized within the bulk solution generating temporal oscillations of the microtubular concentration of the whole solution or within distances of the order of 1 mm allowing to form propagating waves. This may be mediated by millimeter scale variations in the tubulin composition of the medium. The range of the clouds of tubulin-GDP produced by dense nodes of microtubules should control the distance at which the reactivity of 'neighboring' microtubules can be modified and the microtubules to be synchronized. The present study only reports how tubulin-GDP heterodimers diffuse from simplified microtubules. Further advances including reactions based on rescue-catastrophe dynamics \citep{Sur:S01,Ned:JCB02,Ned:COCB03} or more accurate kinetics such as those described in \citet{Van:BJ05} will allow understanding, at the microscopic level, what controls microtubule synchronization over mesoscopic to macroscopic distances.
\\
\\
\indent In this article, further than the study realized on microtubule self-organization, we addressed the question of the robustness of a microscopic process (\textit{i.e.} here, the theoretical communication between microtubules by way of tubulin trails) that, confronted to ambient noise or accidents, could drive the elements of the system to the appearance of macroscopic order. In biology, physics or biocomputing, more and more studies deal now with this question where the dynamical self-organization of a system depends on collective dynamics based on microscopic processes of communication, often in a noisy context (\textit{i.e.} thermal agitation, external perturbations ...) \citep{Les:BR08}. If Nature works since billions years to produce and select processes that are robust in such perturbed contexts for realizing a coherent {--} macroscopic {--} function that plays a role in living organisms, this is not the case for our engineered products. Now, theoreticians and engineers expect from nano or microscopic natural {--} biological, physical {--} systems or from systems inspired from nature (\textit{e.g.} synthetic fiber-shape assemblies \citet{Rot:JACS04,Rot:PLOS04,Gla:IJUC08}) to realize tasks or computations. This knowledge presents a real interest in collision based computing \citep{BLC:CSF05,Iga:LNCS06} or in dynamic self-organizing molecular processors \citep{Pfa:B00,teuscher:026106,Gla:MCBSCG09} in which the wires are not as clearly defined as in electronic processors or brains \citep{Dem:AR01} but constituted by the agents themselves and by their pathways \citep{Gla:MCBSCG09}. The spatial scale level is certainly the most critical criterion that controls the efficiency of these self-ordered microscopic processes to realize the expected macroscopic actions \citep{Con:B95} or to generate the expected processing architectures \citep{Pfa:B00,teuscher:026106,Gla:MCBSCG09}. Studies of how information or matter is exchanged between agents (as presented here or in \citet{Liz:PRE08,Gla:MCBSCG09} of similar systems are a necessary preliminary to such unconventional computing or robotic works, but also concern strongly all the agent-based models of molecular and supramolecular organization in cellular biology, a field that tends to replace (or at least complement) increasingly the differential equation based models. Our work, although not in favor of a reaction-diffusion based computation with
microtubules, could be used in such a way, for determining the efficiency of information or matter exchange in similar systems such as actin comets or other potential trail systems, \textit{i.e.} chemotactic cells \citep{Bag:TH06}.
%
%
%
\begin{acknowledgements}\label{acknowledgements}
The author want to thank O. Bastien (PCV, CEA Grenoble), J. Berro (University of Lyon) and Vic Norris (University of Rouen) for their help and stimulating discussions.
\end{acknowledgements}
%
%
\bibliographystyle{spbasic}      
\bibliography{Glade_Tubulin_Trails}
%

\end{document}